\documentclass[lettersize,journal]{IEEEtran}
\usepackage{amsmath,amsfonts}
\usepackage{algorithmic}
\usepackage{algorithm}
\usepackage{array}
\usepackage{textcomp}
\usepackage{stfloats}
\usepackage{url}
\usepackage{verbatim}
\usepackage{graphicx}
\usepackage{cite}

\usepackage{algorithmic}
\usepackage{textcomp}
\usepackage{color, soul}
\usepackage{tabu}                     
\usepackage{multicol}                 
\usepackage{multirow}                
\usepackage{float}                    
\usepackage{makecell}                 
\usepackage{booktabs}                 
\usepackage{colortbl}
\usepackage{pifont} 

\usepackage{amssymb}
\usepackage{graphicx}
\usepackage{mathrsfs}
\usepackage{amsmath}
\usepackage{makecell}                 
\usepackage{graphicx}
\usepackage{caption} 
\usepackage{subcaption}
\usepackage{ulem}

\usepackage[utf8]{inputenc} 
\usepackage[T1]{fontenc}    
\usepackage{hyperref}       
\usepackage{url}            
\usepackage{booktabs}       
\usepackage{amsfonts}       
\usepackage{nicefrac}       
\usepackage{microtype}      
\usepackage{xcolor}         

\begin{document}

\title{ESG-Net: Event-Aware Semantic Guided Network for Dense Audio-Visual Event Localization}

\author{Huilai Li, 
Yonghao Dang, 
Ying Xing, 
Yiming Wang, 
Jianqin Yin*, \IEEEmembership{Member, IEEE}
\thanks{This work was supported by State Grid Corporation Science and Technology Project Funding (Project Code: 5700-202422243A-1-1-ZN), the National Natural Science Foundation of China (Grant No. 62173045), the Beijing Natural Science Foundation under Grant F2024203115, and the China Postdoctoral Science Foundation under Grant Number 2024M750255.}
\thanks{Huilai Li, Ying Xing, Yiming Wang and Jianqin Yin* are with the School of Intelligent Engineering and Automation, Beijing University of Posts and Telecommunications, Beijing 100876, China (e-mail: lihuilai@bupt.edu.cn; 	xingying@bupt.edu.cn; ymwang99@bupt.edu.cn; jqyin@bupt.edu.cn) (Corresponding author: Jianqin Yin).}
\thanks{Yonghao Dang is with the School of Artificial Intelligence, Beijing University of Posts and Telecommunications, Beijing 100876, China (e-mail: dyh2018@bupt.edu.cn).}
}

\markboth{Journal of IEEE Transactions on Multimedia}%
{Shell \MakeLowercase{\textit{et al.}}: A Sample Article Using IEEEtran.cls for IEEE Journals}


\maketitle

\begin{abstract}
Dense audio-visual event localization (DAVE) aims to identify event categories and locate the temporal boundaries in untrimmed videos. For such challenging task settings, most studies only employ audio-visual event semantic constraints on the final outputs, lacking progressive cross-modal semantic bridging in intermediate layers. This causes semantic gaps that hinder alignment between representations in audio and visual features, making it difficult to distinguish between event-related and irrelevant background content. Moreover, they rarely consider the correlations between events, which limits the model to infer co-occurring events among complex scenarios. In this paper, we incorporate multi-stage semantic guidance and multi-event relationship modeling, which respectively enable progressive semantic understanding of audio-visual events and adaptive extraction of event dependencies, thereby better focusing on event-related information. Specifically, our event-aware semantic guided network (ESG-Net) includes a early semantic interaction (ESI) module and a mixture of dependency experts (MoDE) module. ESI applys multi-stage semantic guidance to explicitly constrain the model in learning semantic information through multi-stage feature fusion and several classification loss functions, ensuring multi-stage understanding of event-related content. MoDE promotes the extraction of multi-event dependencies through multiple serial mixture of experts with adaptive weight allocation. Extensive experiments demonstrate that our method significantly surpasses the state-of-the-art methods, while greatly reducing parameters and computational load. Our code will be released on \url{https://github.com/uchiha99999/ESG-Net}.
\end{abstract}


\section{Introduction}
\IEEEPARstart{D}{ense} audio-visual event localization (DAVE) \cite{geng2023dense} is proposed for detecting events that are both visible and audible in untrimmed videos, while also regressing the temporal boundaries for each event. DAVE allows multiple overlapping events in a long video which is practical for real-life use cases such as surveillance systems \cite{elharrouss2021review}, sports analytics \cite{sha2018interactive} and video event tagging \cite{wang2012event}. Due to the challenging setting of this task, how to sufficiently handle multi-modal information to focus on event-related content is crucial for DAVE. 

\begin{figure*}[t]
	\centering
	\includegraphics[width=18cm,height=5.8cm]{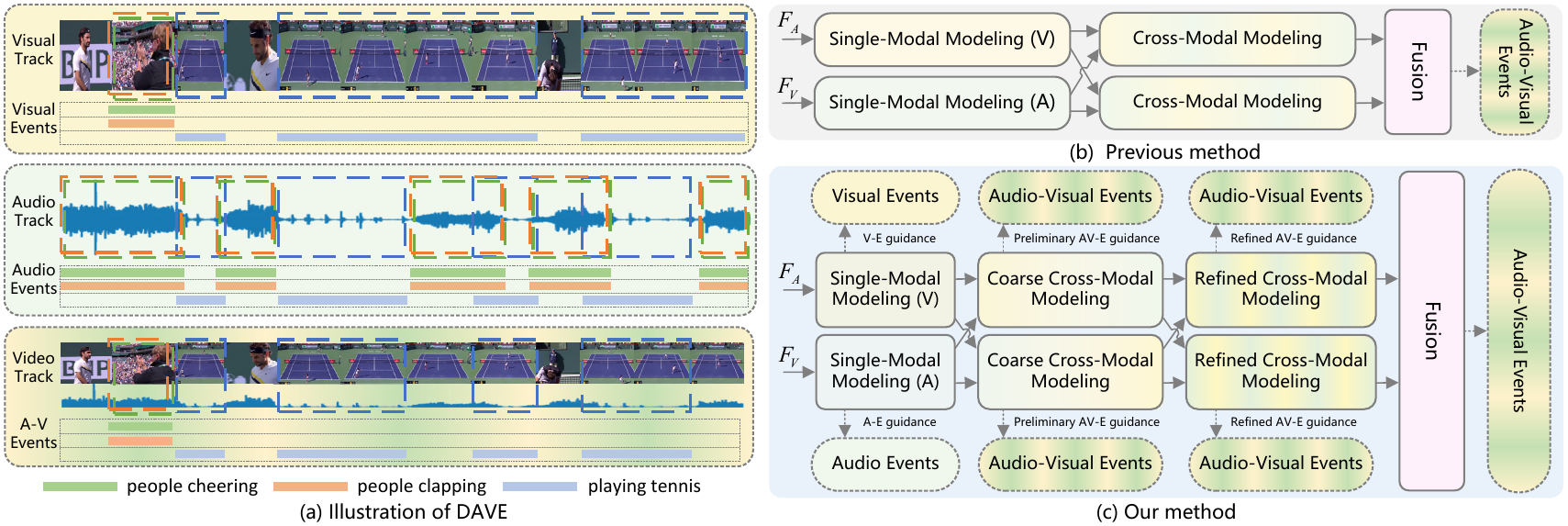}
	\caption{(a) Audio-visual events is intersection of visual and audio events, while significant differences exist between visual and audio content. (b) Most previous methods only performe semantic constraints on the final outputs. (c) We consider event-related semantic bridging between different modalities in intermediate layers through multi-stage fusion and semantic guidance.}
	\label{fig_intro}
\end{figure*}

In recent years, most existing cross-modal frameworks for detecting dense events only perform event-related constraints (event classification and boundary localization losses) on the final outputs \cite{geng2023dense, geng2024uniav}. While other methods \cite{xing2024locality} apply contrastive learning \cite{khosla2020supervised} for cross-modal alignment in intermediate layers at the feature level. However, the lack of semantic constraints during multi-modal feature fusion makes localizing audio-visual events in untrimmed videos particularly challenging. This is because the inherent temporal misalignment between audio and visual modalities creates significant semantic gaps (e.g., a visual scene of playing tennis accompanied by delayed cheers in Figure~\ref{fig_intro}). Insufficient event supervision and forced alignment of multi-modal features without semantic constraints can introduce much noise due to mismatched video segments, making it difficult for the model to distinguish between event-related and irrelevant background content. Therefore, it is crucial to impose sufficient cross-modal semantic bridging in intermediate layers to achieve semantic alignment for accurate audio-visual event location. 

Additionally, few studies take the correlations between audio-visual events into consideration, which limits the model's ability to infer co-occurring events in complex scenarios. Specifically, there are complex dependencies between different audio-visual events \cite{geng2023dense} (\emph{e.g.}, thunder is often accompanied by rain). These dependencies are variable (\emph{e.g.}, piano solo or playing together with guitar) and may differ in order and overlap duration. Approaches with fixed modeling structures are insufficient to fit such diverse dependency patterns. How to adaptively capture those dependencies in complex scenarios to improve the reasoning capability of the model for co-occurring events is also a crucial issue for DAVE.

In this work, we propose an event-aware semantic guided network (ESG-Net), which leverages multi-stage semantic guidance for progressive audio-visual events understanding and multi-event relationship modeling for adaptive dependencies capture, respectively. \textbf{(1)} Inspired by audio-visual video parsing (AVVP) \cite{tian2020unified, lin2021exploring, mo2022multi, rachavarapu2023boosting}, our model is designed to output event detection results from single-modal stage to multi-modal stages with consistent semantic supervision. As shown in Figure~\ref{fig_intro}, compared with previous methods, ESG-Net can achieve cross-modal semantic bridging in intermediate layers through multi-stage constraints, effectively reducing the multi-modal semantic gap and guiding the model to focus on foreground information (\emph{e.g.}, audio-visual events like "playing tennis" tagged in Figure~\ref{fig_intro}). \textbf{(2)} Additionally, we also consider the dependencies among multiple events and adaptively capture these complex multi-event relationships by mixture of experts (MoE) \cite{masoudnia2014mixture, gormley2019mixture, zhou2022mixture} module. Overall, our method progressively emphasizes event-related content through multi-stage processing, where the semantic bridging and dependency modeling jointly enable accurate audio-visual event detection.

Specifically, ESG-Net consists of an early semantic interaction (ESI) module and a mixture of dependency experts (MoDE) module. \textbf{(1)} The ESI module includes multi-stage feature fusion and multi-stage semantic guidance. We progressively fuse multi-modal information and perform cross-modal semantic bridging by weighted classification loss functions across three stages: single-modal attention, audio(visual)-driven mixture, cross-modal pyramid. These stages respectively enable single-modal event detection, preliminary audio-visual event detection, and refined audio-visual event detection, ensuring sufficient semantic alignment between audio and visual branches in intermediate layers. \textbf{(2)} MoDE is proposed to address the the modeling of correlations between audio-visual events with efficient computation. MoDE consists of multiple sequential mixture of experts \cite{masoudnia2014mixture, gormley2019mixture, zhou2022mixture} layers, with each layer selecting only one expert at a time. Through flexible selection of experts in different layers, diverse event dependencies can be captured. 

Our event-aware semantic guided network can gradually focus on audio-visual content across multiple intermediate layers and capture multi-event relationships for accurate dense events localization. The results show that ESG-Net surpasses the baseline model by 2.1\% on I3D+VGGish \cite{carreira2017quo, hershey2017cnn} backbone and 4.1\% on ONE-PEACE \cite{wang2023one} backbone. Our contributions are summarized as: 
\begin{itemize}
\item We propose an effective, efficient and adaptive framework, ESG-Net, for dense audio-visual events location in untrimmed videos, which can pay more attention to event-related information in complex scenarios.
\item We introduce an early semantic interaction module and a mixture of dependency experts module, designed for cross-modal bridging by explicit semantic guidance in intermediate layers and adaptive multi-event dependencies capture, respectively. 
\item Abundant experiments demonstrate that our event-aware semantic guided network outperforms state-of-the-art methods across various intersection-over-union thresholds.
\end{itemize}

\section{Related Work}
\label{Related}

\subsection{Audio-Visual Event Localization}
The audio-visual event localization (AVEL) task was first proposed by Tian et al. \cite{tian2018audio}, who also introduced the AVE dataset and a dual multi-modal residual network. Then some studies \cite{wu2019dual, duan2021audio} provided supervised cross-modal fusion and event detection methods, while others \cite{xu2020cross, xuan2020cross, zhou2021positive, xia2022cross} are compatible with both weakly-supervised and supervised methods. Recently, Zhou et al. \cite{zhou2025towards} proposed an open-vocabulary benchmark to address the closed-set limitation in AVEL, which enables recognition of unseen events. However, these studies are based on trimmed video clips, with a single event in each clip, which are not fit with the real-world scenarios. To address this issue, Geng et al. \cite{geng2023dense} released UnAV-100, a real-world dataset containing a large number of untrimmed videos with dense events, and introduced a benchmark for DAVE. While Krishna et al. \cite{ranjay2017dense} also proposed a similar task called dense video captioning (DVC), which simultaneously detects events and generates textual descriptions. 

Localizing audio-visual events from untrimmed videos is a challenging setting, requiring models to simultaneously achieve multi-modal fusion and parse video segments. Vladimir et al. \cite{iashin2005better, iashin2020multi} and Xie et al. \cite{xie2023global} respectively achieved multi-modal DVC by integrating audio representations and globally-shared text representations with visual track. Geng et al. \cite{geng2024uniav} subsequently proposed a unified framework to integrate three tasks, including temporal action localization (TAL) \cite{lin2019bmn, liu2022end, zhang2022actionformer}, sound event detection (SED) \cite{kim2022sound, xiao2023fmsg}, and AVEL \cite{tian2018audio, wu2019dual, jiang2023leveraging}. Faegheh et al. \cite{sardari2025coleaf} improved the weakly-supervised AVVP \cite{tian2020unified, lin2021exploring, mo2022multi, rachavarapu2023boosting} by employing a reference and an anchor module to extract single-modal and multi-modal information. Recently, Tang et al. \cite{tang2024avicuna} upgraded Vicuna to AVicuna, enhancing the multi-modal understanding capabilities of large language models by aligning audio and video features with text. Xing et al. \cite{xing2024locality} proposed a locality aware cross-modal correspondence learning framework and improved the inaccurate recognition of event boundaries \cite{geng2023dense} by using contrastive learning and dynamic local temporal attention. While Zhou et al. \cite{zhou2024dense} improve the performance through adaptive cross-modal attention and temporal modeling at different granularities. However, these methods only employ event-related constraints on the final outputs, lacking cross-modal semantic bridging in intermediate layers, which remain modality semantic gap between audio and visual features. In this work, we perform multi-stage semantic guidance for multi-modal alignment, which allows the model to focus on event-related content in multiple intermediate layers. 

\subsection{Audio-Visual Bridging}
The joint analysis of audio and video is a powerful tool that can be applied to various tasks \cite{shahabaz2024increasing}. How to reasonably bridge different modalities is crucial for models to understand task-related content. Early researches \cite{arandjelovic2017look, owens2018audio, kikuchi2018watch} simply concatenated and projected the audio and visual features extracted by sub-networks. This may result in insufficient features fusion and suboptimal extraction of task-specific information. Subsequently, researchers used the audio-visual similarity to guide attention toward relevant regions for particular tasks \cite{tian2018audio, chung2019perfect, hu2019deep, wu2019dual, wang2020alignnet, lee2021looking}. Some methods also applied contrastive learning for self-supervised synchronous tasks \cite{chung2017out, korbar2018cooperative, sung2023sound}. Currently, there are typical frameworks mainly include single-modal modeling and multi-modal modeling, which are used for audio-visual correspondence (AVC) \cite{majumder2024learning}, audio-visual parsing (AVP) \cite{xu2024rethink}, and AVEL \cite{xuan2020cross, xu2020cross, mo2022multi, geng2023dense, he2024cace}. Most of these methods only empoly late multi-modal fusion and perform semantic constraints on the final outputs, ignoring semantic cross-modal bridging in intermediate layers, which may make the model difficult to focus on task-related content as early as possible. 

Other studies attempt to use early cross-modal bridging to reduce modality gap. Introducing contrastive learning \cite{khosla2020supervised} is a reasonable method to strengthen audio-visual connections. For example, Hong et al. \cite{hong2023hyperbolic} proposed hyperbolic alignment to implement zero-shot learning. Trevine et al. \cite{oorloff2024avff} achieved video deepfake detection by aligning the multi-modal features of frames. Lee et al. \cite{lee2022sound} conducted pairwise contrastive learning to map audio, images, and text into a multi-modal shared latent space, while Mo et al. \cite{mo2023unified} built a framework for localization, separation, and recognition by matching positive samples. Although these methods can effectively optimize multi-modal features, there are many mismatched segments in most untrimmed videos. Forcibly performing frame-level alignment may introduce significant noise. In addition to contrastive learning, some studies used co-attention to capture the relationships between multi-modal inputs \cite{afouras2020self, duan2021audio, zhou2021positive, xia2022cross, geng2023dense}. Other methods also added a parallel early fusion branch alongside two single-modal branches \cite{hu2016temporal, ramaswamy2020see, hu2023cross} for temporal alignment. However, these audio-visual bridging methods ignore the semantic alignment between different modalities in intermediate processes. Our method employs a multi-stage guided classification to enforce explicit cross-modal semantic bridging gradually, thereby reducing modality semantic gap and promoting the attention of our model to task-related content.

\section{Method}
\label{Method}

\begin{figure*}[t]
	\centering
	\includegraphics[width=18cm,height=9.56cm]{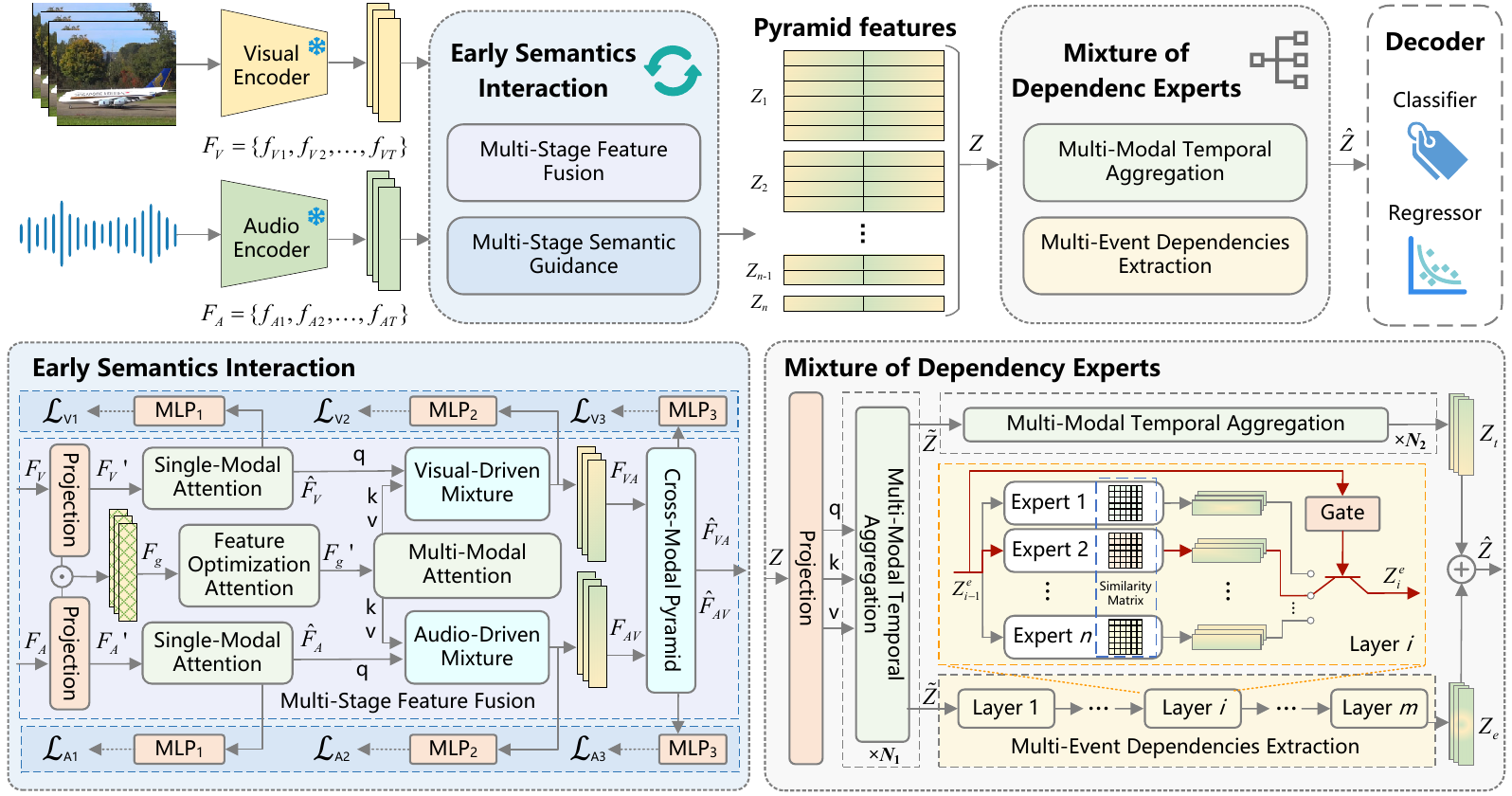}
	\caption{Overview of ESG-Net. The original audio and video are fed into frozen encoders for feature extraction. Then, the ESI perform cross-modal early fusion and multi-stage semantic guidance to focus on event-related content in intermediate layers. Subsequently, the MoDE, which consists of multiple mixture of experts layers, is used to extract mutil-event dependencies from the integrated features. Finally, the audio-visual features are decoded to obtain event categories and temporal boundaries.}
	\label{fig1}
\end{figure*}

To overcome the multi-modal semantic gap between audio and visual features in intermediate layers and the modeling of event correlations for DAVE, we design an early semantic interaction module to achieve multi-stage cross-modal semantic bridging, and a mixture of dependency experts module to capture complex relationships between events. An overview of the event-aware semantic guided network we propose is shown in Figure~\ref{fig1}.

\subsection{Problem Setting}
The aim of DAVE is to recognize the categories of all events occurring in the given untrimmed videos and to locate the temporal boundaries of each event. Formally, the inputs consist of $T$ visual snippets $V=\{V_{t}\}_{t=1}^{T}$ and $T$ audio snippets $A=\{A_{t}\}_{t=1}^{T}$, where $T$ is determined by the length of the video. For each visual-audio snippet, the groundtruth is represented as $Y=\{c_{n}, t^{s}_{n}, t^{e}_{n}\}_{n=1}^{N}$, where $N$ is the number of events, $t^{s}_{n} $ and $ t^{e}_{n}$ are the start and end time of $n$-th event, and $c_{n}$ is the event category. Similarly, the prediction result of the model is $\hat{P}=\{\hat{c}_{n}, \hat{t}^{s}_{n}, \hat{t}^{e}_{n}\}_{n=1}^{N}$. Note that $\{t^{s}_{n}, t^{e}_{n}\}$ and $\{\hat{t}^{s}_{n}, \hat{t}^{e}_{n}\}$ are class-aware, existing only when $c_{n}\in\{1, \dots, C\}$.

\subsection{Overall Architecture}
ESG-Net comprises four key components: 1) pre-trained audio encoder and visual encoder to extract features from original videos; 2) an early semantic interaction module for cross-modal semantic bridging; 3) a mixture of dependency experts module for capturing multi-event dependencies; 4) a decoder for localizing audio-visual events. These key components are executed in sequence as illustrated in Figure~\ref{fig1}. 

\textbf{Audio Encoder and Visual Encoder.}
Following \cite{geng2023dense, geng2024uniav}, we apply frozen encoder to extract audio features ${F_A} = \{{f_{A1}},{f_{A2}}, \dots ,{f_{AT}}\}$ and visual features ${F_V} = \{{f_{V1}},{f_{V2}}, \dots ,{f_{VT}}\}$. Here, the multi-modal features $\{F_{A}, F_{V}\}\in\mathbb{R}^{T \times D}$ are extracted synchronously, $T$ represents the length of the video, $D$ is the dimension of these features. 

\textbf{Early Semantic Interaction Module.}
To achieve cross-modal semantic bridging in intermediate layers, ESI consists of Multi-Stage Feature Fusion and Multi-Stage Semantic Guidance. The Multi-Stage Feature Fusion includes: Single-Modal Attention for visual/audio modeling, Audio(Visual)-Driven Mixture for preliminary audio-visual modeling, and Cross-Modal Pyramid Module\cite{geng2023dense} for refined audio-visual modeling. The outputs of these three parts are $\{\hat{F}_{A},\hat{F}_{V}\}\in\mathbb{R}^{T_{m} \times D}$, $\{F_{AV},F_{VA}\}\in\mathbb{R}^{T_{m} \times D}$, $\{\hat{F}_{AV},\hat{F}_{VA}\}\in\mathbb{R}^{T_{l}\times D}$, respectively, where $T_{m}$ is the maximum length among all videos, $T_{l}$ is the total length of the multi-level pyramid features. Then, they are used for Multi-Stage Semantic Guidance through classfication loss functions. The details are as shown in Section \ref{ESI}.

\textbf{Mixture of Dependency Experts Module.}
After concatenating $\hat{F}_{AV}$ and $\hat{F}_{VA}$ as $Z\in\mathbb{R}^{T_{l} \times 2D}$, MoDE performs cross-modal temporal aggregation and event dependency modeling. Specifically, MoDE consists of two branches: one for in-depth temporal modeling, and the other revealing multi-event relationships through a cascade of MoE layers. The outputs $Z_{t}\in\mathbb{R}^{T_{l} \times C}$ and $Z_{e}\in\mathbb{R}^{T_{l} \times C}$ from two branches are then added together to obtain $\hat{Z}\in\mathbb{R}^{T_{l} \times C}$ for final decoding. The details of MoDE are as shown in Section \ref{MoDE}.

\textbf{Decoder.}
The decoder $\mathscr{D}$, following \cite{geng2023dense, geng2024uniav}, includes a classification head and a regression head. Each of them consists of three layers of 1D convolutions and one layer of activation function (Sigmoid for the classification head, ReLU for the regression head). The integrated features $\hat{Z}$ across all pyramid levels are fed into $\mathscr{D}$ to predict the probability $p(c_{n})$ of event categories and the distances $D_{e}=\{d^{s}_{n}, d^{e}_{n}\}_{n=1}^{N}$ from the current moment $t$ to the temporal boundaries of events $T_{e}=\{t^{s}_{n}, t^{e}_{n}\}_{n=1}^{N}$. Note that the regression head here is class-aware, its outputs are available only when audio-visual events are detected.

\subsection{Early Semantic Interaction}
\label{ESI}

Due to the lack of cross-modal semantic bridging in intermediate layers, the modality semantic gap caused by temporal misalignment between audio and video lays obstacles to the understanding of audio-visual events in previous methods \cite{geng2023dense, geng2024uniav}. To solve this issue, we propose Multi-Stage Feature Fusion and Multi-Stage Semantic Guidance in ESI that enable progressive understanding of audio-visual events. The former enables the model to fuse audio-visual information gradually, while the latter performs semantic bridging for progressive event understanding. The details are as shown in the lower left part of Figure~\ref{fig1}:

\textbf{Multi-Stage Feature Fusion.}
Before cross-modal semantic bridging, we need to fuse multi-modal information gradually so that the model can better reduce modality gaps among intermediate layers. As shown in Figure~\ref{fig1}, our multi-stage feature fusion can be divided into \emph{Single-Modal Attention}, \emph{Audio(Visual)-Driven Mixture} and \emph{Cross-Modal Pyramid}, which are used for single-modal temporal modeling, multi-modal information fusion , and multiple temporal resolution modeling, respectively. 

\emph{(1) Single-Modal Attention.}
To capture long-term temporal relations among single-modal segments and filter out some noise, we perform single-modal temporal modeling. Following this stage, both the audio and visual modalities are able to preliminarily identify the approximate locations of events. Specifically, we first optimize the original features through projection layers, where $F_{A}'=f_{proj}(F_{A})$ and $F_{V}'=f_{proj}(F_{V})$. Then the mapped audio and visual features $\{F_{A}',F_{V}'\}\in\mathbb{R}^{T \times D}$ are used for temporal modeling through single-modal transformer blocks. Each single-modal transformer block includes a multi-headed self-attention (MSA) and a feed-forward network (FFN), which can be denote as:
\begin{equation}
\label{ASMA}
\hat{F_{A}}=\text{FFN}(\text{MSA}(F_{A}'W_{q}^{A},F_{A}'W_{k}^{A},F_{A}'W_{v}^{A}))
\end{equation}
\begin{equation}
\label{VSMA}
\hat{F_{V}}=\text{FFN}(\text{MSA}(F_{V}'W_{q}^{V},F_{V}'W_{k}^{V},F_{V}'W_{v}^{V}))
\end{equation}
where $\{\hat{F_{A}},\hat{F_{V}}\}\in\mathbb{R}^{T_{m} \times D}$ are the outputs of audio and visual features, respectively. $W_{q}^{A}$, $W_{k}^{A}$, $W_{v}^{A}$, $W_{q}^{V}$, $W_{k}^{V}, W_{v}^{V}$$\in\mathbb{R}^{D \times D}$ are learnable parameters. 

\emph{(2) Audio(Visual)-Driven Mixture.}
After single-modal attention, we attempt to coarsely fuse audio and visual information to achieve preliminary perception of audio-visual events. Compared with direct interaction with another modality, our audio(visual)-driven mixture enables single-modal features to acquire complementary information from early multi-modal representations with smaller modality gaps (because early multi-modal representations retain some correlations and synchronized temporal structures). 

We first build the early multi-modal representations, which contain frame-aligned audio-visual information. However, due to the significant differences between original audio and visual features $\{F_{A},F_{V}\}$, simply concatenating them to get the early multi-modal representations may introduce much noise and lead to insufficient fusion. So we first design an aligner that includes two projection layers and a feature optimization attention block to map them onto the same representation space: 
\begin{equation}
\label{project}
F_{g}=f_{proj}(F_{A}) \odot f_{proj}(F_{V})
\end{equation}
\begin{equation}
\label{optimize}
F_{g}'=\text{MSA}(F_{g}W_{q}^{g},F_{g}W_{k}^{g},F_{g}W_{v}^{g})
\end{equation}
where $F_{g}'\in\mathbb{R}^{T_{m} \times D}$ represents the early multi-modal representation, $W_{q}^{g}$, $W_{k}^{g}$, $W_{v}^{g}$$\in\mathbb{R}^{T_{m} \times T_{m}}$. 
Then, like the \emph{Single-Modal Attention} in Eq. (\ref{ASMA}) and Eq. (\ref{VSMA}), the early multi-modal representation $F_{g}'$ is also employed for temporal modeling:
\begin{equation}
\hat{F_{g}}=\text{FFN}(\text{MSA}(F_{g}'W_{q}^{g'},F_{g}'W_{k}^{g'},F_{g}'W_{v}^{g'}))
\end{equation}
where $\hat{F_{g}}\in\mathbb{R}^{T_{m} \times D}$, $W_{q}^{g'}$, $W_{k}^{g'}, W_{v}^{g'}$$\in\mathbb{R}^{D \times D}$. Now the early multi-modal representation $\hat{F_{g}}$ contains frame-aligned audio and visual information. We introduce an audio(visual)-driven mixture to extract multi-modal complementary information for $\{\hat{F_{A}},\hat{F_{V}}\}$. The audio(visual)-driven mixture includes a multi-headed cross-attention (MCA) and a FFN, as follows: 
\begin{equation}
F_{AV}=\text{FFN}(\text{MCA}(\hat{F_{g}}W_{q}^{g_{1}},\hat{F}_{A}W_{k}^{A},\hat{F}_{A}W_{v}^{A}))
\end{equation}
\begin{equation}
F_{VA}=\text{FFN}(\text{MCA}(\hat{F_{g}}W_{q}^{g_{2}},\hat{F}_{V}W_{k}^{V},\hat{F}_{V}W_{v}^{V}))
\end{equation}
where $W_{q}^{g_{1}}, W_{k}^{A}, W_{v}^{A}, W_{q}^{g_{2}}, W_{k}^{V}, W_{v}^{V}\in\mathbb{R}^{D \times D_{m}}$. $\hat{F_{g}}$ provides the query vectors, while $\hat{F}_{A}$ and $\hat{F}_{V}$ provide the key and value vectors.

\emph{(3) Cross-Modal Pyramid.}
Our ESI are designed to facilitate progressive cross-modal interaction while gradually reducing modality gaps between two branches, thereby enabling sufficient feature fusion. After the single-modal attention and the audio(visual)-driven mixture, our model achieves audio(visual) temporal aggregation and preliminary cross-modal interaction. To further reduce the modality gap and endow our model with fine-grained temporal boundary awareness (events vary in duration from 0.2s to 70s), we perform directly cross-modal fusion at different temporal resolutions following \cite{geng2023dense}. Specifically, we temporally downsample the features $\{F_{AV},F_{VA}\}$ with the stride $2^{l_{c}-1}$, where $l_{c}$ is the index of the current pyramid block. There is a cross-attention after a downsampling of audio and visual features. The outputs of the cross-modal pyramid module are $\{\hat{F}_{AV},\hat{F}_{VA}\}\in\mathbb{R}^{T_{l}\times D}$, where $T_{l}$ is the total length of the multi-level pyramid features.

\textbf{Multi-Stage Semantic Guidance.}
In this section, we aim to enhance audio-visual event localization by enabling the model to understand event-related content in intermediate layers. However, even with multi-stage feature fusion and temporal modeling, the model still struggles to reduce the semantic gap between audio and visual branches. That is, without explicit semantic constraints in intermediate layers, the model cannot understand and locate event-related segments at early stages. Inspired by \cite{tian2020unified, qian2020multiple}, we apply an event-related semantic guidance that directly facilitate audio-visual events understanding through multi-stage constraints. Specifically, we employ MLPs on the outputs of \emph{Multi-Stage Feature Fusion} ($\{\hat{F}_{A},\hat{F}_{V}\}$, $\{F_{AV},F_{VA}\}$, $\{\hat{F}_{AV},\hat{F}_{VA}\}$) for decoding. Each MLP consists of two linear layers with a PReLU activation. Then, we use focal loss functions \cite{ross2017focal} respectively on the outputs after MLPs for semantic consistency constraints. Note that here we only detect the event categories for each snippet, without regressing the temporal boundaries. The total multi-stage classification loss $\mathcal{L}_{mcls}$ is as:

\begin{equation}
\label{mcls}
\mathcal{L}_{mcls}=\sum_{i=1}^{n}\alpha_{i}(\mathcal{L}_{Ai}+\mathcal{L}_{Vi})
\end{equation}
where $i\in\{1,2,3\}$, $\alpha_{i}$ is the weight of losses. $\alpha_{1}$, $\alpha_{2}$ and $\alpha_{3}$ increase sequentially to ensure the effectiveness of multi-stage optimization from single-modal to multi-modal information. Through these stages of progressive semantic constraints in intermediate layers, \emph{Single-Modal Attention} can coarsely detect audio and visual events individually, while \emph{Audio(Visual)-Driven Mixture} and \emph{Cross-Modal Pyramid} can achieve preliminary and further refined audio-visual event detection, respectively.

\subsection{Mixture of Dependency Experts}
\label{MoDE}

The mixture of dependency experts module is designed for final multi-modal fusion and multi-event correlations extraction. As shown in the lower right part of Figure~\ref{fig1}, MoDE mainly includes a Multi-Modal Temporal Aggregation module and a Multi-Event Dependencies Extraction module. 

\textbf{Multi-Modal Temporal Aggregation.}
Before localizing audio-visual events, we need to aggregate the results of ESI $\{\hat{F}_{AV},\hat{F}_{VA}\}$ and apply temporal modeling to obtain a comprehensive multi-modal representation. By doing this, the model can capture cross-modal long-term relations and filter out irrelevant noise. Specifically, $\hat{F}_{AV}$ and $\hat{F}_{VA}$ are first concatenated together as $Z$. Then we input it into a projection layer to get $Z'\in\mathbb{R}^{T_{l} \times C}$ for preliminary fusion while filtering out redundant information, where $C$ is dimension of $Z'$, and it is also the number of event categories. Then, to further refine multi-modal information, we employ a temporal modeling for $Z'$ as in Eq. (\ref{ASMA}) and Eq. (\ref{VSMA}), composed of $N_{1}+N_{2}$ multi-headed self-attention. $\tilde{Z}\in\mathbb{R}^{T_{l} \times C}$is the output after $N_{1}$ temporal attention blocks, and $Z_{t}$ is the final output after $N_{1}+N_{2}$ temporal attention blocks. 

\textbf{Multi-Event Dependencies Extraction.}
In audio-visual scenarios, there are dependencies between co-occurring events (\emph{e.g.}, thunder is often accompanied by rain and wind). However, these correlations are complex and variable across different scenarios, and the events may have different degrees of overlap. We hope that the model can reveal event dependencies automatically according to different inputs. Mixture-of-Experts model \cite{masoudnia2014mixture, gormley2019mixture, zhou2022mixture}, as an approach capable of automatically routing, are well-suited for adaptive complex event dependency processing. Inspired by \cite{yang2021two}, we further employ a series of MoE layers to handle such issue as shown in the lower right part of Figure~\ref{fig1}. Each MoE layer consists of $n$ experts, with only one expert selected at a time. By combining different experts across $m$ layers, complex multi-event dependencies can be revealed adaptively (each MoE layer has $n$ experts, there would be $n^{m}$ possible combinations). Specifically, to ensure that only one expert is selected in each layer at a time, we designed the MoE gate with a Gumbel-Softmax \cite{jang2016categorical}. The output of $i$-th layer is:
\begin{equation}
Z_{i}^{e}=Gate_{i}\times\text{stack}(E_{i})
\end{equation}
\begin{equation}
Gate_{i}=\text{GumbelSoftmax}(f_{MLP}(Z_{e}^{i-1}))
\end{equation}
where $E_{i}=\{E_{i,1},E_{i,2},\dots ,E_{i,n}\}$ is the experts set of $i$-th MoE layer. The input to the first layer is the mid-process output $\tilde{Z}$ mentioned in Multi-Modal Temporal Aggregation. Each expert in $E_{i}$ is composed of a convolutional layer, an adjacency matrix $\mathcal{A}\in\mathbb{R}^{C \times C}$ and a LeakyReLU activation as : 
\begin{equation}
E_{i,j}=\text{LeakyReLU}(\mathcal{A}\times \text{Convolution}(Z_{e}^{i-1}))
\end{equation}
where $E_{i,j}$ is the $j$-th expert in the $i$-th layer, $Z_{e}^{i-1}\in\mathbb{R}^{T_{l} \times C}$ is the output of $(i-1)$-th MoE layer, $\mathcal{A}$ is initialized as a learnable unit diagonal matrix and adapts to uncover latent dependencies, $C$ is the number of event categories. Finally, the output of the last MoE layer $Z_{e}$ is added with $Z_{t}$ to obtain $\hat{Z}$ for decoding.

\subsection{Training and Inference}
\textbf{Training.} We employ an end-to-end training approach, where three losses are used to constrain the following aspects: a focal loss \cite{ross2017focal} $\mathcal{L}_{cls}$ for event classification (while addressing class imbalance by focusing on hard examples), a generalized IoU loss \cite{rezatofighi2019generalized} $\mathcal{L}_{reg}$ for temporal boundary estimation (while providing scale-invariant regularization and stable gradients for non-overlapping intervals), and $\mathcal{L}_{mcls}$ in Section \ref{ESI}. The loss functions are expressed as:
\begin{equation}
\label{totalloss}
\mathcal{L}=\mathcal{L}_{cls}+\mathcal{L}_{reg}+\mathcal{L}_{mcls}
\end{equation}
$\mathcal{L}_{cls}$ and $\mathcal{L}_{reg}$ are used to constrain the final audio-visual event detection and regression, while $\mathcal{L}_{mcls}$ is used for cross-modal semantic bridging in intermediate layers. More details are provided in the supplementary materials.

\textbf{Inference.} When given all snippets of a video during inference, the model will output candidates that include event categories and temporal boundaries, each accompanied by a confidence score. Then, all candidates are processed by multi-class Soft-NMS \cite{bodla2017soft} to filter out temporal boundaries with high overlap within the same event category.

\section{Experiments}
\label{Experiments}

\subsection{Experimental Settings}
\label{settings}
\textbf{Datasets.} 
The UnAV-100 dataset proposed by Geng et al. \cite{geng2023dense} consists of 10,790 videos with a total duration exceeding 126 hours, and includes 30,059 audio-visual events across 100 event categories. In addition to the raw videos, UnAV-100 also provides audio and video features extracted by frozen encoders. It includes two versions: (1) The optical flow features are extracted by RAFT \cite{teed2020raft}, the RGB features are extracted by two-stream I3D \cite{carreira2017quo}, while audio features are extracted by VGGish \cite{hershey2017cnn}; (2) Audio and visual features are extracted by ONE-PEACE \cite{wang2023one}, where the visual encoder is fine-tuned on Kinetics-400 \cite{kay2017kinetics}. As the standard dataset distribution, the training, validation and testing set are split at a ratio of 3:1:1.

\textbf{Evaluation Metrics.} 
We use the same split as in \cite{geng2023dense} for training, validation, and testing. For the evaluation of event detection tasks with temporal localization, we used the mean Average Precision (mAP) at the tIoU thresholds [0.5:0.1:0.9] (\emph{e.g.}, between 0.5 and 0.9, with a step size of 0.1) and the average mAP at the tIoU thresholds [0.1:0.1:0.9].

\begin{table}[t]
  \centering
  \caption{\textbf{Comparison results with previous methods.} The audio features are extracted by I3D \cite{carreira2017quo}, the visual features are extracted by VGGish \cite{hershey2017cnn}. Mean Average Precision at the tIoU thresholds [0.5:0.1:0.9] are reported.}
    \begin{tabular}{c|cccccc}
    \toprule
    Method & 0.5   & 0.6   & 0.7   & 0.8   & 0.9   & Avg. \\
    \midrule
    VSGN \cite{zhao2021video}  & 24.5  & 20.2  & 15.9  & 11.4  & 6.8   & 24.1  \\
    TadTR \cite{liu2022end} & 30.4  & 27.1  & 23.3  & 19.4  & 14.3  & 29.4  \\
    ActionFormer \cite{zhang2022actionformer} & 43.5  & 39.4  & 33.4  & 27.3  & 17.9  & 42.2  \\
    TriDet \cite{shi2023tridet} & 46.2  & -     & -     & -     & -     & 44.4  \\
    UnAV \cite{geng2023dense}  & 50.6  & 45.8  & 39.8  & 32.4  & 21.1  & 47.8  \\
    UniAV(AT) \cite{geng2024uniav} & 49.3  & -     & -     & -     & -     & 47.0  \\
    UniAV(STF) \cite{geng2024uniav} & 50.1  & -     & -     & -     & -     & 48.2  \\
    CCNet \cite{zhou2024dense} & 51.9  & 47.2  & 41.5  & \textbf{34.1} & \textbf{23.0} & 49.2  \\
    LoCo \cite{xing2024locality}  & 52.8  & 47.6  & 41.1  & 33.3  & 21.9  & 49.5  \\
    \textbf{Ours} & \textbf{53.5} & \textbf{48.2} & \textbf{41.9} & 33.6  & 21.2  & \textbf{49.9} \\
    \bottomrule
    \end{tabular}%
  \label{comp1}%
\end{table}%

\begin{table}[t]
  \centering
  \caption{\textbf{Comparison results with previous methods.} The audio features and visual features are extracted by ONE-PEACE \cite{wang2023one}. Mean Average Precision \ at the tIoU thresholds [0.5:0.1:0.9] are reported.}
    \begin{tabular}{c|cccccc}
    \toprule
    Method & 0.5   & 0.6   & 0.7   & 0.8   & 0.9   & Avg. \\
    \midrule
    ActionFormer \cite{zhang2022actionformer} & 49.2  & -     & -     & -     & -     & 47.0  \\
    TriDet \cite{shi2023tridet} & 49.7  & -     & -     & -     & -     & 47.3  \\
    UnAV \cite{geng2023dense}  & 53.8  & 48.7  & 42.2  & 33.8  & 20.4  & 51.0  \\
    UniAV(AT) \cite{geng2024uniav} & 54.1  & 48.6  & 42.1  & 34.3  & 20.5  & 50.7  \\
    UniAV(STF) \cite{geng2024uniav} & 54.8  & 49.4  & 43.2  & 35.3  & 22.5  & 51.7  \\
    CCNet \cite{zhou2024dense} & 57.3  & 52.2  & 46.2  & \textbf{38.1} & 25.6  & 54.1  \\
    LoCo \cite{xing2024locality}  & 56.5  & 51.4  & 44.7  & 36.7  & \textbf{25.9} & 53.4  \\
    \textbf{Ours} & \textbf{58.7} & \textbf{53.4} & \textbf{46.7} & 38.0  & 24.0  & \textbf{55.1} \\
    \bottomrule
    \end{tabular}%
  \label{comp2}%
\end{table}%

\subsection{Comparison with Existing Work}
\label{comparison}
As shown in Table~\ref{comp1} and Table~\ref{comp2}, we conduct comprehensive performance comparisons between the proposed method and existing methods. These results show that ESG-Net outperforms other methods in overall performance when using I3D+VGGish \cite{carreira2017quo, hershey2017cnn} or ONE-PEACE \cite{wang2023one} for feature extracting. Compared to the existing state-of-the-art TAL models (VSGN \cite{zhao2021video}, TadTR \cite{liu2022end}, ActionFormer \cite{zhang2022actionformer} and TridDet \cite{shi2023tridet}), ESG-Net demonstrates significant performance improvements at any threshold. When compared to the baseline method (UnAV \cite{geng2023dense}), our approach still shows improvements of 2.1\% and 4.1\% under the two above feature extractors. While compared to the state-of-the-art DAVE models, our method only underperforms LoCo \cite{xing2024locality} or CCNet \cite{zhou2024dense} at high tIoU thresholds (0.8 and 0.9), but our overall performance still surpasses that of these models. Combining the experimental results from two tables, our method achieves the highest average mAP under two versions of feature extractors, demonstrating its strong adaptability and robustness. 

It is noteworthy that in Table~\ref{comp1} and Table~\ref{comp2}, our method performs excellently at low tIoU thresholds (0.5, 0.6, and 0.7), but it is slightly weaker than CCNet \cite{zhou2024dense} and LoCo \cite{xing2024locality} at tIoU thresholds of 0.8 and 0.9. From the perspective of model structure: (1) CCNet \cite{zhou2024dense} mainly performs adaptively temporal modeling at different resolutions, enabling it to focus on information at various temporal granularities; (2) LoCo \cite{xing2024locality} employs dynamic attention to adjust the temporal scale of the features, which solves the issue of asynchronization in dense audio-visual events detection. In summary, CCNet \cite{zhou2024dense} and LoCo \cite{xing2024locality} directly enhance the accuracy by imposing strong constraints on the temporal boundaries of audio-visual events. Although our method also achieves significant improvements over the baseline methods, these gains are mainly attributed to the enhanced capability of detecting audio-visual events (thereby mitigating the miss detection). However, since our cross-modal semantic bridging applies audio-visual event constraints instead of audio/visual event constraints in single-modal branches, it inevitably introduces some noise (\emph{e.g.}, the audio-visual event occurs in 3s-5s, the corresponding audio event actually occurs in 2s-6s). Therefore, our method achieves better audio-visual event detection at low tIoU, while exhibiting suboptimal performance at higher tIoU due to the semantic constraints with temporal noise. This can also explain why $\alpha_{i}$ in Eq. (\ref{mcls}) needs to increase sequentially (noise in multi-stage semantic constraints is reduced sequentially). In summary, CCNet \cite{zhou2024dense} and LoCo \cite{xing2024locality} pay more attention on temporal boundaries of audio-visual events, with their performance being effective at high tIoU thresholds, whereas our method is more sensitive to the events themselves, thus achieving better performance at low tIoU thresholds.

\begin{table*}[t]
  \centering
  \caption{\textbf{Pseudo-label accurancy in AVVP task.} }
    \begin{tabular}{c|ccccc|ccccc|c}
    \toprule
    \multirow{2}[4]{*}{Method} & \multicolumn{5}{c|}{Segment-level}    & \multicolumn{5}{c|}{Event-level}      & \multirow{2}[4]{*}{Avg.} \\
\cmidrule{2-11}          & A     & V     & AV    & Type  & Event & A     & V     & AV    & Type  & Event &  \\
    \midrule
    HAN+PPL \cite{rachavarapu2024weakly} & 62.5  & 55.3  & 52.3  & 56.0  & 58.3  & 55.4  & 51.1  & 46.9  & 50.9  & 50.6  & 53.9  \\
    MA+PPL \cite{rachavarapu2024weakly} & 61.7  & 61.8  & 57.5  & 60.6  & 59.4  & 55.4  & 57.9  & 51.6  & 55.0  & 52.6  & 57.4  \\
	 VALOR \cite{lai2023modality} & \textbf{80.4} & 71.8  & 63.7  & \ul{72.0}  & \textbf{79.7} & \ul{72.2}  & 65.9  & 55.6  & 64.6  & 68.0  & 69.4  \\
    UWAV \cite{lai2025uwav}  & 78.4  & \textbf{74.5} & \ul{65.5}  & \textbf{72.8} & 78.4  & 71.1  & \textbf{69.6} & \ul{57.7}  & \ul{66.1}  & \ul{69.0}  & \ul{70.3}  \\
    UWAV+ESI & \ul{79.2}  & \ul{73.6}  & \textbf{65.8} & \textbf{72.8} & \ul{78.8}  & \textbf{72.7} & \ul{68.9}  & \textbf{59.5} & \textbf{67.0} & \textbf{69.9} & \textbf{70.8} \\
    \bottomrule
    \end{tabular}%
  \label{avvp}%
\end{table*}

\begin{table}[t]
  \centering
  \caption{\textbf{The ablation study of ESI.} $\mathcal{L}_{AVi}$ denotes semantic guided classification loss function in each stage as Eq. (\ref{mcls}), 'ER' denotes the early multi-modal representations. The best mAP in each column is bolded while the second-best is underlined. Mean Average Precision at the tIoU thresholds [0.5:0.1:0.9] are reported.}
    \resizebox{\columnwidth}{!}{
    \begin{tabular}{cccc|cccccc}
    \toprule
    $\mathcal{L}_{AV1}$ & $\mathcal{L}_{AV2}$ & $\mathcal{L}_{AV3}$ & ER  & 0.5   & 0.6   & 0.7   & 0.8   & 0.9   & Avg. \\
    \midrule
          & $\checkmark$   & $\checkmark$   & $\checkmark$    & 58.4  & 52.9  & 46.1  & \ul{37.4} & 23.2  & \ul{54.7}  \\
    $\checkmark$   &       & $\checkmark$   & $\checkmark$    & \ul{58.5}  & \textbf{53.5} & \ul{46.5} & 36.9  & 23.0  & 54.6  \\
    $\checkmark$   & $\checkmark$   &       & $\checkmark$   & 54.9  & 49.3  & 43.0  & 35.1  & \textbf{24.1}  & 51.8  \\
    $\checkmark$   &       &       &      & 55.0  & 49.6  & 43.8  & 35.3  & 22.7 & 51.6  \\
       &    $\checkmark$   &       &      & 55.7  & 50.0  & 43.5  & 35.6  & 22.6 & 52.2  \\
       &       &   $\checkmark$    &      & 57.0  & 51.0  & 44.2  & 35.7  & 22.0 & 53.3  \\
          &       &       & $\checkmark$  & 54.7  & 49.8  & 43.3  & 36.2  & \ul{24.0}  & 51.6  \\
    $\checkmark$   & $\checkmark$   & $\checkmark$   &       & 57.6  & 51.6  & 44.5  & 35.6  & 21.8  & 53.5  \\
          &       &       &       & 55.2  & 49.9  & 43.8  & 35.0  & 22.0  & 51.6  \\
    $\checkmark$   & $\checkmark$   & $\checkmark$   & $\checkmark$   & \textbf{58.7} & \ul{53.4}  & \textbf{46.7}  & \textbf{38.0}  & \ul{24.0}  & \textbf{55.1} \\
    \bottomrule
    \end{tabular}}
  \label{abla1}%
\end{table}%

\begin{table}[t]
  \centering
  \caption{\textbf{The ablation study of different structure in ESI.} The best mAP in each column is bolded. Mean Average Precision at the tIoU thresholds [0.5:0.1:0.9] are reported.}
    \begin{tabular}{c|cccccc}
    \toprule
    Method & 0.5   & 0.6   & 0.7   & 0.8   & 0.9   & Avg. \\
    \midrule
    Cat   & 55.1  & 49.5  & 43.4  & 34.6  & 22.6  & 51.6  \\
    CMPTE(UnAV) & 54.8  & 49.4  & 42.9  & 35.1  & 22.9  & 51.3  \\
    CMCC(CCNet) & 54.1  & 48.8  & 42.6  & 34.5  & 22.6  & 50.8  \\
    Ours  & \textbf{58.7}  & \textbf{53.4}  & \textbf{46.7} & \textbf{38.0} & \textbf{24.0} & \textbf{55.1} \\
    \bottomrule
    \end{tabular}
  \label{abla12}
\end{table}

\begin{table}[t]
  \centering
  \caption{\textbf{The ablation study of different structure in MoDE.} The best mAP in each column is bolded while the second-best is underlined. Mean Average Precision at the tIoU thresholds [0.5:0.1:0.9] are reported.}
    \begin{tabular}{c|cccccc}
    \toprule
    Method & 0.5   & 0.6   & 0.7   & 0.8   & 0.9   & Avg. \\
    \midrule
    MBM   & 58.4  & 53.0  & 45.7  & 37.0  & 22.9  & \ul{54.5}  \\
    GCN   & 58.4  & 52.6  & 46.1  & 36.8  & 22.4  & 54.2  \\
    Attention & 58.0  & 52.6  & 45.9  & 37.1  & \ul{23.9}  & 54.4  \\
	 MoS$^2$ Block & \ul{58.7}  & 53.2  & 46.3  & 37.1  & 22.0  & 54.7  \\
    TDM(UnAV) & \textbf{59.2} & \textbf{53.5} & \ul{46.5}  & \ul{37.4}  & 23.7  & \textbf{55.0} \\
    Ours  & \ul{58.7}  & \ul{53.4}  & \textbf{46.7} & \textbf{38.0} & \textbf{24.0} & \textbf{55.1} \\
    \bottomrule
    \end{tabular}
  \label{abla22}
\end{table}

\begin{table}[t]
  \centering
  \caption{\textbf{The ablation study on the number of experts in each MoDE layer.} Mean Average Precision at the tIoU thresholds [0.5:0.1:0.9] are reported.}
    \begin{tabular}{c|rrrrrr}
    \toprule
    Num of Experts & \multicolumn{1}{c}{0.5} & \multicolumn{1}{c}{0.6} & \multicolumn{1}{c}{0.7} & \multicolumn{1}{c}{0.8} & \multicolumn{1}{c}{0.9} & \multicolumn{1}{c}{Avg.} \\
    \midrule
    0     & 58.1  & 52.7  & 46.0  & 37.2 & 21.9  & 54.3  \\
    1     & 58.4  & 52.6  & 46.1  & 36.8  & 22.4  & 54.2  \\
    \textbf{2} & \textbf{58.7} & \textbf{53.4} & \textbf{46.7} & \textbf{38.0}  & \textbf{24.0}  & \textbf{55.1} \\
    4     & 57.7  & 51.8  & 44.4  & 36.8  & 23.4 & 53.6  \\
    8     & 58.2  & 52.7  & 45.5  & 36.2  & 22.4  & 54.2  \\
    \bottomrule
    \end{tabular}%
  \label{abla3}%
\end{table}%

\begin{table}[t]
  \centering
  \caption{\textbf{The ablation study on the layers of MoE in our MoDE module.} 'Inf.(sec)' denotes total inference time in seconds. The best mAP in each column is bolded while the second-best is underlined. Mean Average Precision at the tIoU thresholds [0.5:0.1:0.9] are reported.}
   \resizebox{\columnwidth}{!}{
    \begin{tabular}{c|cccccc|c}
    \toprule
    Layers & 0.5   & 0.6   & 0.7   & 0.8   & 0.9   & Avg.  & Inf.(sec) \\
    \midrule
    2     & \ul{58.5}  & 53.2  & 46.3  & 37.1  & 23.0  & 54.7  & \textbf{155.7} \\
    4     & \textbf{58.7}  & \textbf{53.4}  & \textbf{46.7}  & \textbf{38.0}  & \textbf{24.0}  & \textbf{55.1}  & \ul{161.0} \\
    8     & 57.8  & 52.5  & 46.2  & 36.7  & \ul{23.5}  & 54.3  & 172.8 \\
    16    & \textbf{58.7} & \ul{53.3} & \ul{46.5} & \ul{37.5} & \textbf{24.0} & \ul{54.9} & 193.3 \\
    32    & 57.9  & 52.3  & 45.4  & 36.7  & \ul{23.5}  & 54.2  & 239.5 \\
    \bottomrule
    \end{tabular}}
  \label{abla4}%
\end{table}%

\subsection{Comparison of Pseudo-Label Accuracy on AVVP}
We also extend the multi-modal semantic guidance in intermediate layers of ESI to a similar task, audio-visual video parsing. The experimental results are shown in Table~\ref{avvp}. UWAV \cite{lai2025uwav} is pre-trained on the UnAV-100 dataset and then evaluated for pseudo-label accuracy on the Look, Listen, and Parse (LLP) dataset \cite{tian2020unified}. We compute macro F1-scores for three event types: audio (A), visual (V), and audio-visual (AV). Type@AV is the mean of the F1-scores for the A, V, and AV events, while Event@AV is the F1-score of all events in any modality. By integrating multi-stage semantic guidance, "UWAV+ESI" establishes semantic bridging between audio and video modalities within intermediate layers. The results show that "UWAV+ESI" achieves state-of-the-art performance in pseudo-label accuracy on the LLP test set, particularly for audio-visual events, demonstrating both effectiveness and generalizability of ESI.

\subsection{Ablation Study}
To analyze the impact of different components on our model in detail, we conduct ablation experiments based on features extracted by ONE-PEACE \cite{wang2023one}.

\textbf{Key Components in ESI.} As shown in Table~\ref{abla1}, we present the impact of each component on the performance within our ESI module. Firstly, as shown in the first three rows of Table~\ref{abla1}, we remove $\mathcal{L}_{AV1}$, $\mathcal{L}_{AV2}$, and $\mathcal{L}_{AV3}$, respectively, to analyze the role of different component of multi-stage semantic guidance. It can be seen that $\mathcal{L}_{AV3}$ has the most significant impact on performance, while $\mathcal{L}_{AV1}$ and $\mathcal{L}_{AV2}$ have some impact on performance. The results from the fourth to sixth rows in this table exhibit the same pattern. These results indicate that by employing multi-stage semantic guidance for cross-modal semantic bridging, the capability of our model to localize audio-visual events can be enhanced.

Secondly, the results from the fourth to sixth rows in Table~\ref{abla1} individually demonstrate the contribution of $\mathcal{L}_{AVi}$ in ESI, while the last four rows demonstrate the contribution of $\mathcal{L}_{mcls}$ and the early fusion branch to our model: (1)Without early multi-modal representations, $\mathcal{L}_{AVi}$ can bring some improvement for event detection, but the performance remains limited by the feature diffierence between two single-modal branches; (2) When only the early multi-modal representations are used, there is almost no improvement on the performance compared to the original backbone \cite{geng2023dense}; (3) When only the multi-stage semantic guidance ($\mathcal{L}_{mcls}$) is used, the model's performance shows a significant improvement; (4) When both the early multi-modal representations and $\mathcal{L}_{mcls}$ are used, the model achieves the highest average mAP, and the results at the tIoU thresholds of [0.5:0.1:0.9] are also the best or the second-best. These phenomena indicate that the performance improvement brought by the early multi-modal representations is not due to an increase in parameters. In fact, it is the collaboration of the multi-modal early fusion with the multi-stage semantic guidance that enables better event-related semantic bridging in intermediate layers.

Finally, we replace the ESI module with alternative architectures, the results are presented in Table~\ref{abla12}. Here, "Cat" denotes using simple concatenation fusion at the early fusion stage; "CMPTE" refers to replacing ESI with the Cross-Modal Pyramid Transformer Encoder from UnAV \cite{geng2023dense}; and "CMCC" indicates substituting ESI with the Cross-Modal Consistency Collaboration module from CCNet \cite{zhou2024dense}. It can be seen that our method achieves significantly superior performance compared to alternative module configurations, with an average improvement of 3.87\% (±0.4\%) while maintaining comparable parameters with these methods. 

\begin{figure*}[t]
  \centering
  \includegraphics[width=14cm,height=4.67cm]{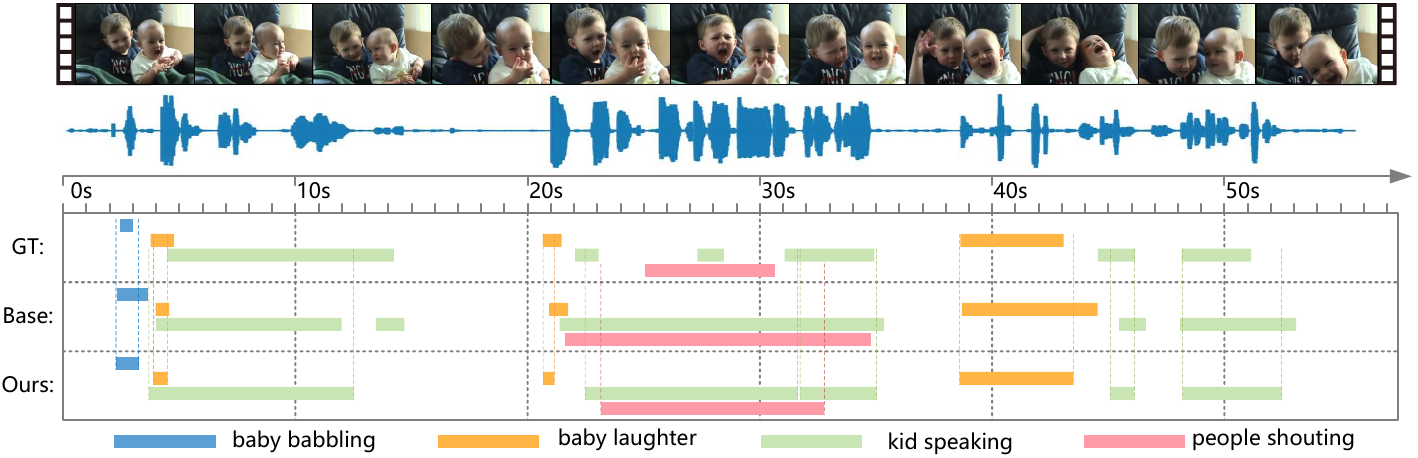}
  \caption{Qualitative results on the testing set. "GT" means ground truth, "Base" denotes the baseline model.}
  \label{qualitative}
\end{figure*}

\textbf{Impact of different structures in MoDE.}
We replace the MoDE module with alternative structures, the results are presented in Table~\ref{abla22}. "MBM" denotes using C*C-matrix-based modeling for event dependencies; "GCN" refers to replacing MoDE with multiple cascaded graph convolutional networks; "Attention" refers to modeling event dependencies by self-attention; "MoS$^2$ Block" refers to adding transformer block after each layer of MoDE and applying expert routing like \cite{jia2024mos2} (we fail to add the shift modules due to dimension mismatch); "TDM" indicates replacing MoDE with the Temporal Dependency Modeling from UnAV \cite{geng2023dense}. Our method achieves the best or the second-best at the tIoU thresholds of [0.5:0.1:0.9] compared with the above structures. Although the average mAP of "TDM" is close to our method, its parameter count exceeds that of MoDE by an order of magnitude. The "MoS$^2$ Block" also adopts a MoE architecture for video processing, but the lack of pre-trained weight results in interference between transformer layers and MoE layers. Finally, compared to static architectures like 'MBM' and 'GCN', MoDE can adaptively uncover relationships among co-occurrring events through dynamic expert selection, demonstrating superior performance. 

\textbf{Impact of the Number of Experts in Each MoDE Layer.} 
We study the impact of the number of experts in each layer of MoDE on the performance, and the results are shown in Table~\ref{abla3}. It can be seen that the mAP reaches the highest value of 55.1\% when the number of experts is set to 2. And the detection accuracy of the model is always the best at the tIoU thresholds of [0.5:0.1:0.9]. However, as the number of experts increases in each layer, the performance of our model gradually declines. Because only one expert can be activated in each layer during one training iteration, adding too many experts in each layer may lead to insufficient training for each expert. 

\textbf{Impact of the MoE Layers in MoDE.} Table~\ref{abla4} shows how the number of MoE layers in MoDE affects the performance. When the number of MoE layers is 16, mAP reaches the second-best at 54.9\%, and when the number of MoE layers is 4, mAP achieves the best at 55.1\%. It can be seen that although increasing the number of MoE layers also increases the total number of experts, the proportion of experts activated during one training iteration remains constant. Increasing the number of MoE layers can achieve more complex combinations of experts as described in Section \ref{MoDE} (each MoE layer has $n$ experts, $m$ MoE layers would have $n^{m}$ possible combinations), which may improve the final performance. However, the training and inference time grows linearly as the number of MoE layers increases. The inference time of 16-layer MoE increases by 20.04\% over 4-layer MoE. After comprehensive consideration of the two metrics, we ultimately adopted the 4-layer MoE which is both effective and efficient.
\begin{figure}[t]
	\centering
	\includegraphics[width=9cm,height=6cm]{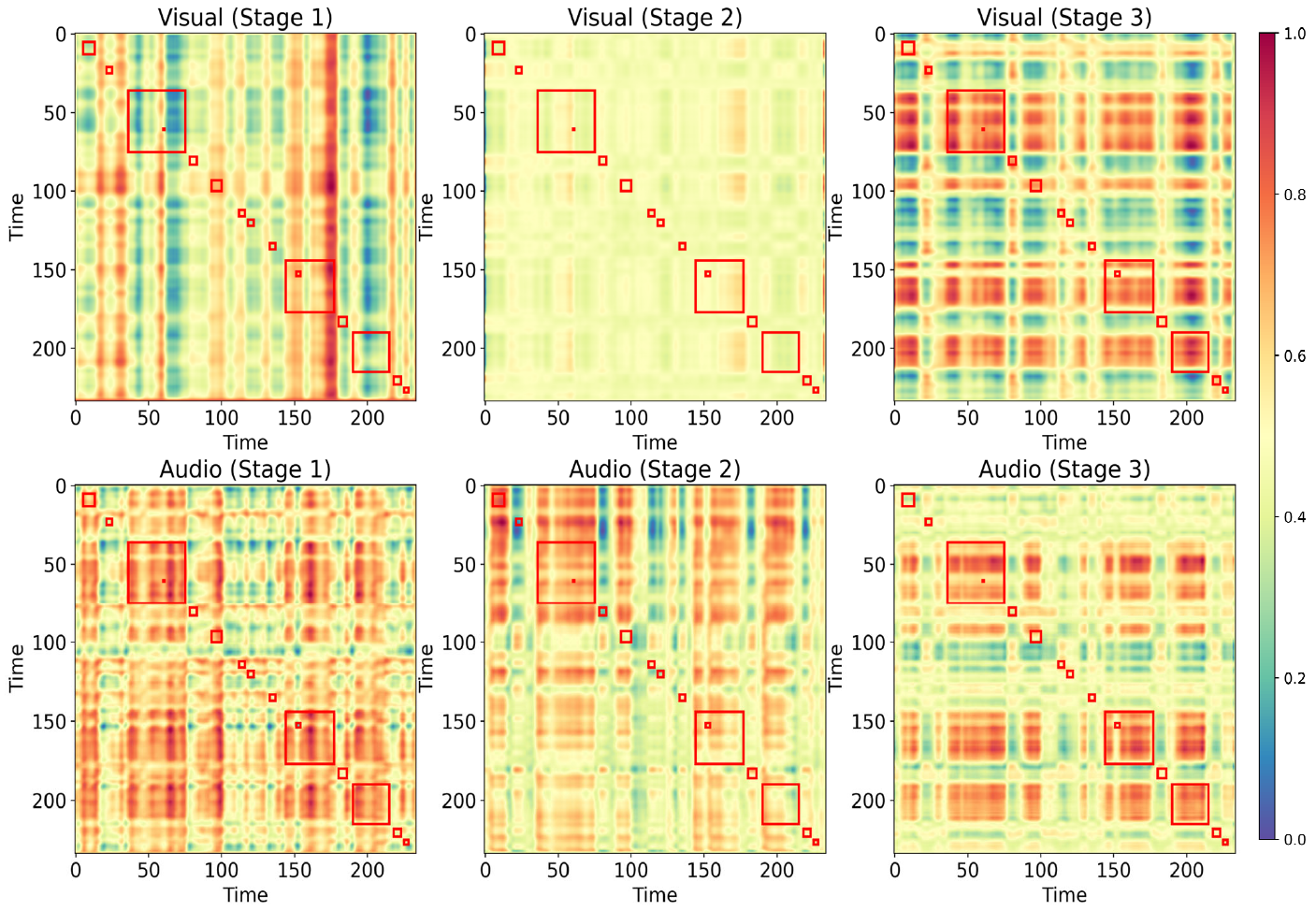}
	\caption{Qualitative results of the impact of multi-stage semantic guidance. Four examples demonstrate the attention of the model to events at different stages. The upper and lower rows in each example display the attention weight matrices of three stages (\emph{Single-Modal Attention}, \emph{Audio(Visual)-Driven Mixture} and \emph{Cross-Modal Pyramid}) for the visual and audio branches. The red boxes indicate the audio-visual events in videos.}
	\label{attmap}
\end{figure}
\begin{figure}[t]
	\centering
	\includegraphics[width=9cm,height=4.756cm]{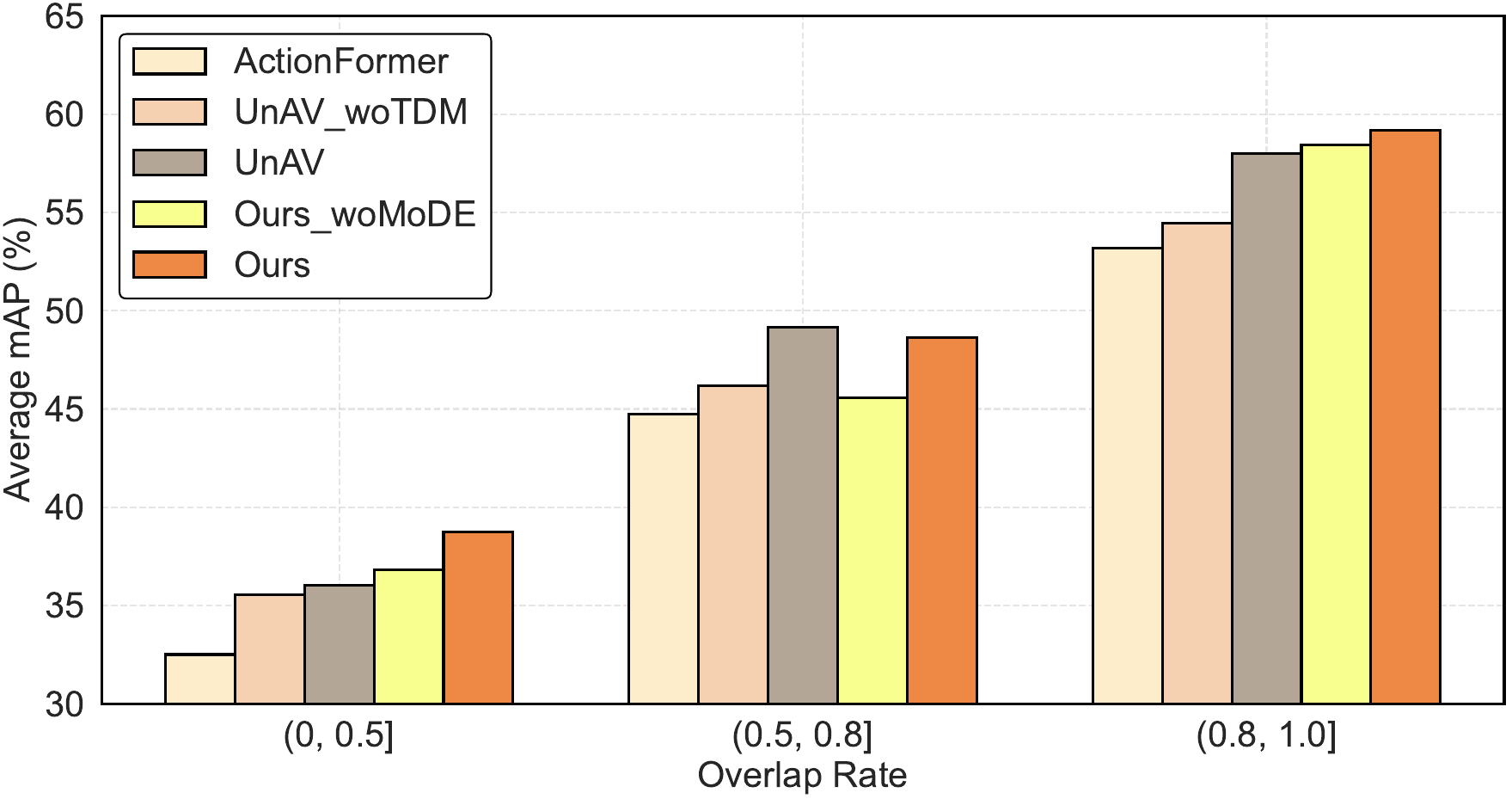}
	\caption{Qualitative results of average performance on co-occurring events with different overlap rates on the videos from the test set.}
	\label{statistics}
\end{figure}
\begin{table}[h]
  \centering
  \caption{\textbf{Comparison of parameters and computational load.} }
    \resizebox{\columnwidth}{!}{
    \begin{tabular}{c|ccccc}
    \toprule
    Method & Params.(M) & FLOPs(G) & Mem.(M)  & Tr. Time(h) & Avg. \\
    \midrule
    UnAV\cite{geng2023dense}  & 140.79  & 60.28  & 3628  & 16.03  & 51.0 \\
    UniAV\cite{geng2024uniav} & 186.00  & 32.83  & 4304  & \textbf{9.04} & 51.7 \\
    LoCo\cite{xing2024locality}  & 103.73  & 31.45  & -  & -   & 53.4 \\
    CCNet\cite{zhou2024dense} & 238.78  & 72.14  & 5790  & 20.31  & 54.1 \\
    \textbf{Ours} & \textbf{86.13} & \textbf{20.36} & \textbf{2617} & 22.50  & \textbf{55.1} \\
    \bottomrule
    \end{tabular}}
  \label{para}
\end{table}

\subsection{Parameter and Computational Load.} To demonstrate the effectiveness and efficiency of our method, we compare the parameter, FLOPs, training time and memory usage of different methods as shown in Table~\ref{para}. Note that we only compare state-of-the-art DAVE methods that are either open-source or well-documented in their papers. It can be seen that our method not only achieves the highest mAP of 55.1\% but also has the lowest consumption in terms of parameter, FLOPs and memory usage. In fact, the main computational consumption of our method is concentrated on ESI, as this module is mainly used for complex temporal modeling and fusion of single-modal and multi-modal features. While MoDE has only about 2M parameters, but it can efficiently model dependencies among multiple events. Secondly, our method exhibits the longest training duration among all compared approaches, which is mainly attributed to the sequential structure within the MoDE module that introduces extended forward propagation pathways. However, the long training duration of CCNet \cite{zhou2024dense} is caused by the scale of its model and high computational demands. In summary, our ESG-Net achieves the best comprehensive performance, significantly reducing computational load while maintaining superior focus on event-related content.

\subsection{Visualization and Discussion}
\textbf{Qualitative Comparison.} We present the qualitative detection results of our method in Figure~\ref{qualitative} and compare with the baseline method \cite{geng2023dense} ("Base"). To better show the performance of ESG-Net and "Base", we add extension lines for each detected event. As we can see, for the events "baby babbling", "baby laughter" and "people shouting", both our method and "Base" are capable of accurately detecting the categories of these events. However, the temporal boundaries regressed by "Base" are often either too early or too late. As for the events "kid speaking", although our model can not accurately output the three sets of temporal boundaries, it still achieved a higher tIoU, especially in the noisy background between 22 seconds and 35 seconds. Overall, these results show the superior performance of detection and boundaries regression of our method. 

\textbf{Impact of Multi-Stage Semantic Guidance.} 
We show the role of our multi-stage semantic guidance by the attention weight matrixs across three stages (\emph{Single-Modal Attention}, \emph{Audio(Visual)-Driven Mixture}, \emph{Cross-Modal Pyramid}) in Figure~\ref{attmap}. The first and second rows of each example are the results of the visual branch and the audio branch, respectively. Each red bounding box denotes a audio-visual event. As we can see, in stage 1, the model does not have a clear focus area, showing lower concern for audio-video events. Additionally, the attention weight matrixs have significant differences between audio and visual branch. In stage 2, our model begins to pay attention to the event-related areas, but their attention to local details remained insufficient. While in stage 3, both the audio branch and the visual branch can basically pay attention to audio-visual events, and the attention weight matrices of the two branches become similar. To sum up, the above results demonstrate that our multi-stage semantic guidance can progressively guide different modalities to focus on event-related information while gradually filtering out irrelevant noise. 

\textbf{Detection capability for co-occurring events.} 
	To evaluate the effectiveness of our MoDE in detecting co-occurring events, we test the mAP on videos that contained co-occurring events with different overlap rates. "UnAV\_woTDM" denotes the UnAV \cite{geng2023dense} without Temporal Dependency Modeling, and "Ours\_woMoDE" represents ESG-Net without MoDE. We conduct experiments using features extracted by I3D+VGGish \cite{carreira2017quo, hershey2017cnn}, with the results shown in Figure~\ref{statistics}. We observed that compared to models that do not consider inter-event modeling ("ActionFormer" \cite{zhang2022actionformer} and "UnAV\_woTDM"), both UnAV \cite{geng2023dense} and our method demonstrate more stable performance when handling the videos with any overlap rates. Compared to UnAV \cite{geng2023dense}, our method achieves better performance on the videos with overlap rates of (0, 0.5] and (0.8, 1.0], demonstrating the effectiveness of MoDE in capturing event dependencies. However, a slight performance degradation of our method is observed on videos with overlap rates in (0.5, 0.8]. This can be attributed to the limited number of test samples within this specific interval, which may lead to bias in the evaluation metrics.

\section{Conclusion}
In this paper, we propose an event-aware semantic guided network that includes an early semantic interaction module and a mixture of dependency experts module for dense audio-visual event localization. ESI introduces the multi-stage feature fusion and implements multi-modal consistency constraints through multi-stage semantic guidance, which achieves cross-modal bridging in intermediate layers and allows model to focus on event-related content. MoDE reveals the multi-event dependencies by adaptively combining suitable experts across different MoE layers. In summary, ESG-Net performs explicit semantic alignment in intermediate layers and multi-event relationships extraction, achieving more accurate detection and localization for audio-visual events. Through detailed ablation experiments, we demonstrate the individual contribution of each component in ESG-Net. Our method not only markedly outperforms state-of-the-art methods in terms of accuracy, but also shows efficient performance in parameters and computational load. 

We end with some limitations and outlining directions for future research. First, we plan to incorporate our method with adaptive temporal attention mechanism and uncertainty-aware modeling to address noisy constraints in single branches. Secondly, current DAVE methods are limited to fixed event categories and lacks detailed textual descriptions of video events. We propose to leverage the pre-trained knowledge in multi-modal large language models by fine-tuning with mixed question-answering data, to develop a unified framework capable of handling temporal-aware audio-visual downstream tasks (\emph{e.g.}, dense audio-visual event localization, video captioning and audio-visual video parsing) in open scenarios.

{
\bibliographystyle{IEEEtran}
\bibliography{ref}
}

\begin{IEEEbiography}[{\includegraphics[width=1in,height=1.25in,clip,keepaspectratio]{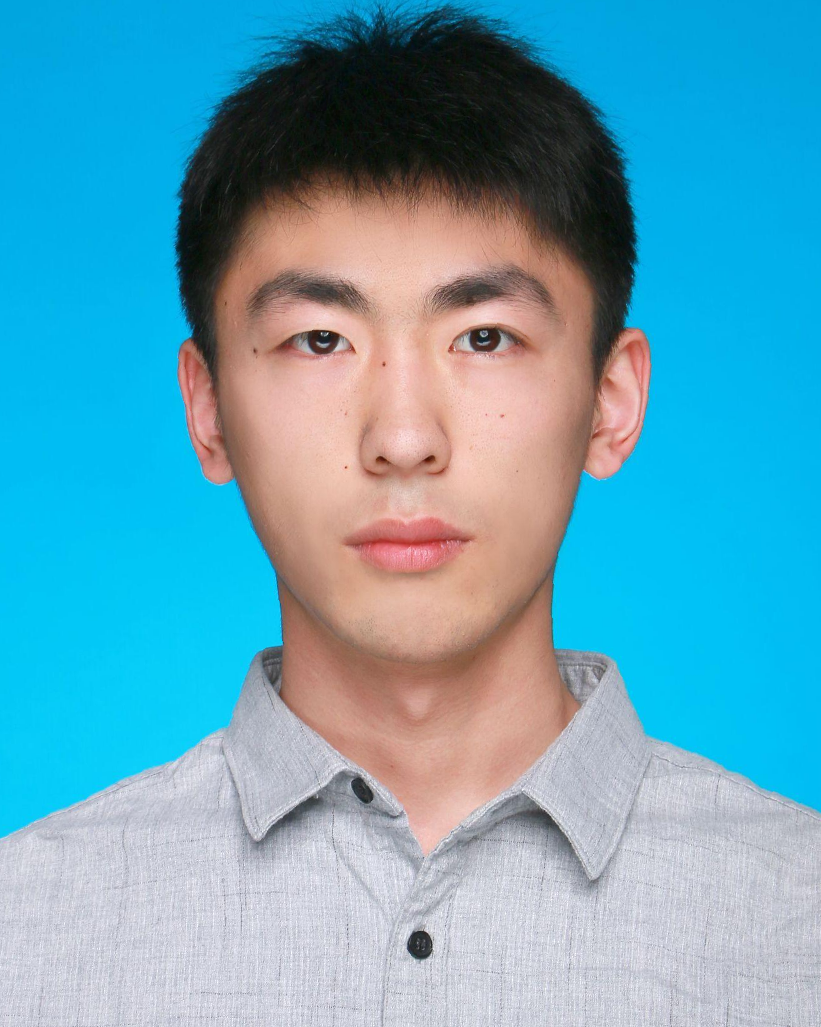}}]{Huilai Li}
received the B.Eng. degree in 2018 from the Zhengzhou University, Zhengzhou, China. He is currently working toward the Ph.D. degree with the School of Intelligent Engineering and Automation, Beijing University of Posts and Telecommunications, Beijing, China. His research interests include multi-modal learning, computer vision, and deep learning.
\end{IEEEbiography}

\begin{IEEEbiography}[{\includegraphics[width=1in,height=1.25in,clip,keepaspectratio]{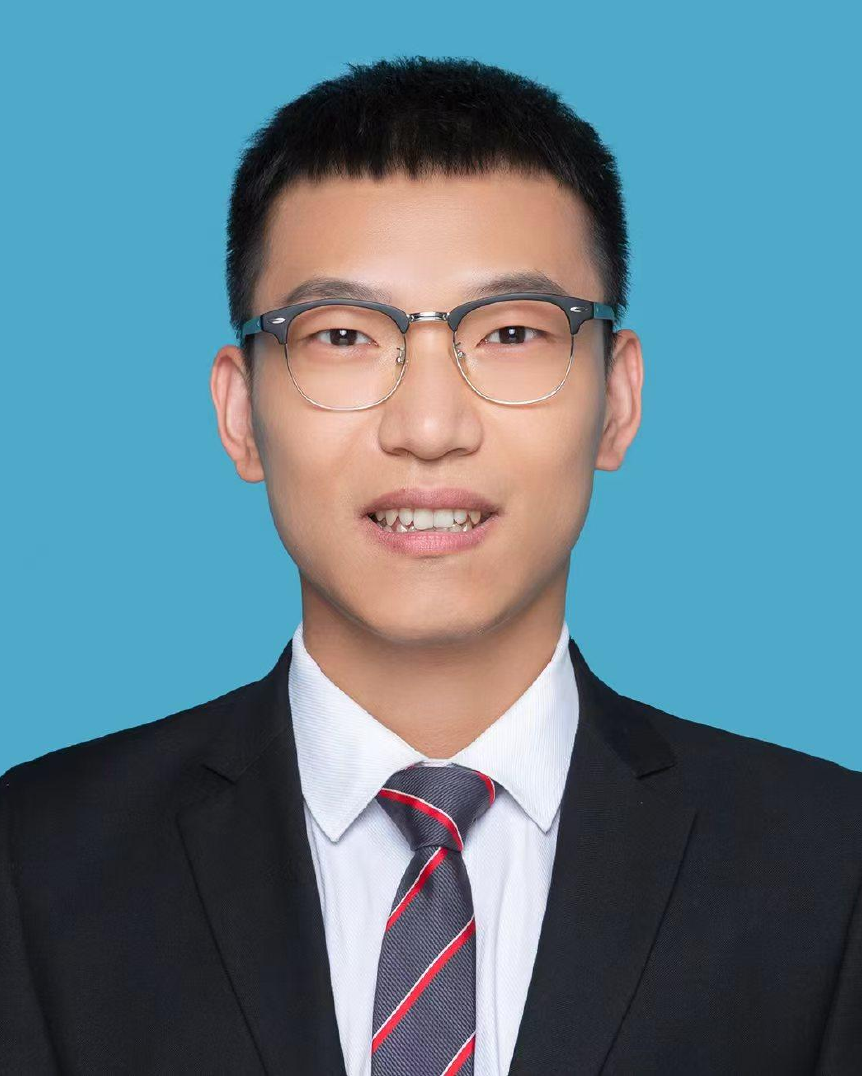}}]{Yonghao Dang}
received Bachelor degree in computer science and technology from the University of Jinan, Jinan, China, in 2018, and the Ph.D. degree from the School of Artificial Intelligence, Beijing
University of Posts and Telecommunications, Beijing, China, in 2023. His research interests include
computer vision, image processing, and deep learning.
\end{IEEEbiography}

\begin{IEEEbiography}[{\includegraphics[width=1in,height=1.25in,clip,keepaspectratio]{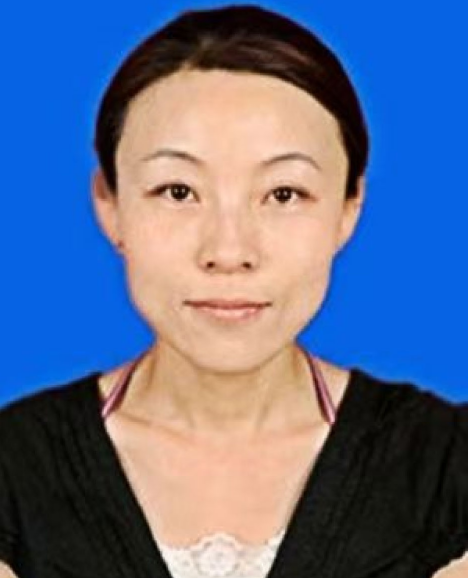}}]{Ying Xing} 
received the Ph.D. degree in Computer Science and Technology from Beijing University of Posts and Telecommunications, Beijing, China in 2014. She is currently an associate professor with the School of Intelligent Engineering and Automation, Beijing University of Posts and Telecommunications. Her research interests include software testing and vulnerability detection.
\end{IEEEbiography}

\begin{IEEEbiography}[{\includegraphics[width=1in,height=1.25in,clip,keepaspectratio]{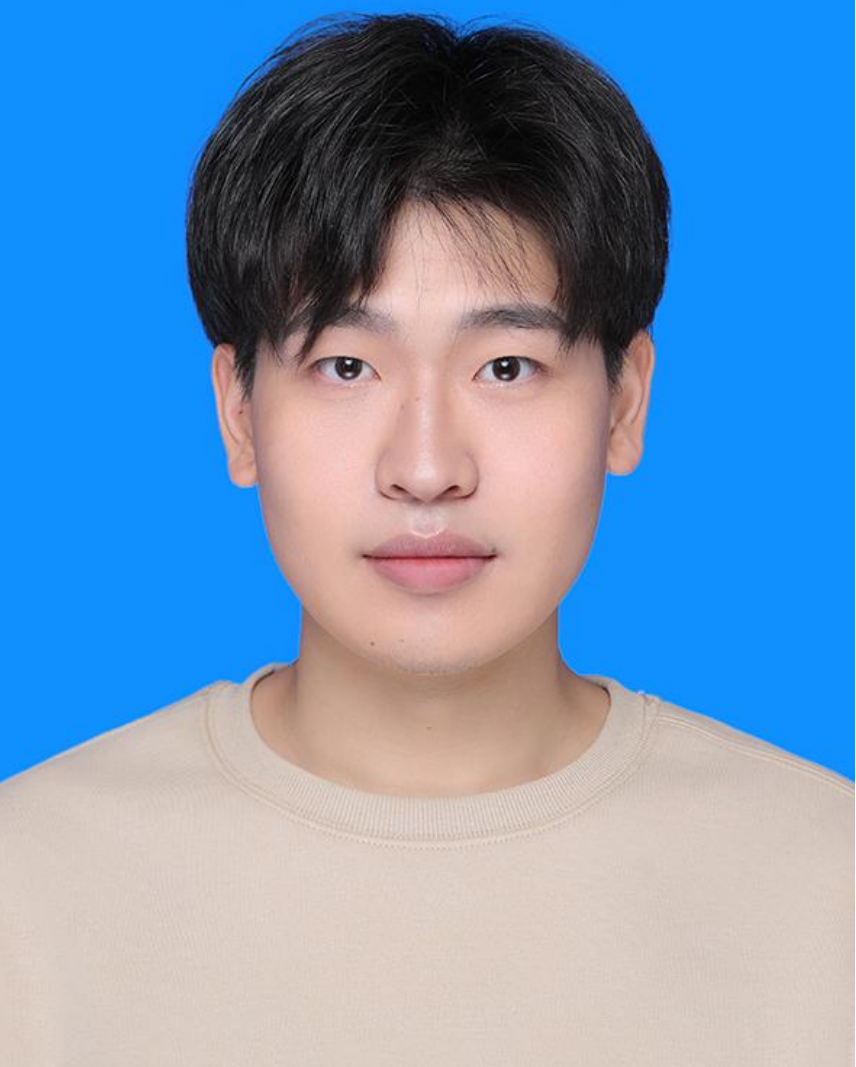}}]{Yiming Wang}
graduated from Hebei GEO University with a Bachelor degree, Shijiazhuang, China. He is currently working toward the Ph.D. degree with the School of Intelligent Engineering and Automation, Beijing University of Posts and Telecommunications, Beijing, China. His research interests include multi-modal learning, computer vision, and deep learning.
\end{IEEEbiography}

\begin{IEEEbiography}[{\includegraphics[width=1in,height=1.25in,clip,keepaspectratio]{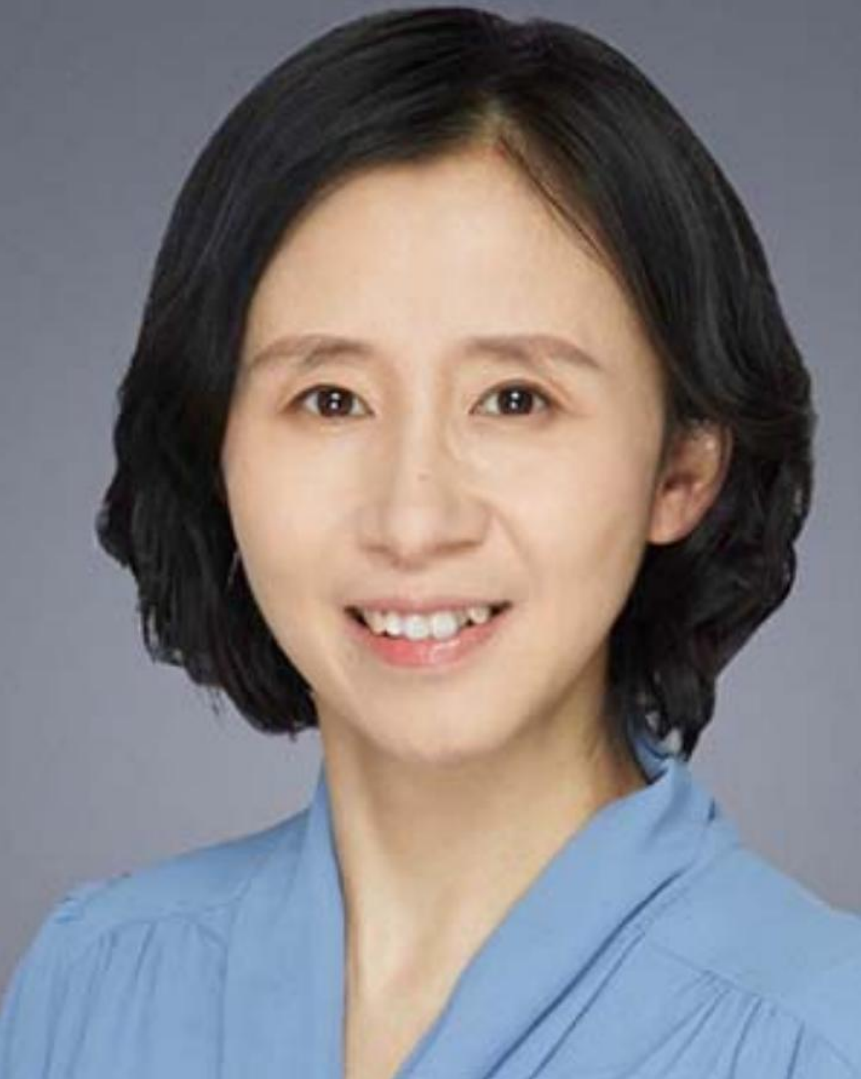}}]{Jianqin Yin}
(Member, IEEE) received the Ph.D. degree from Shandong University, Jinan, China, in 2013. She is currently a Professor with the School of Intelligent Engineering and Automation, Beijing University of Posts and Telecommunications, Beijing, China. Her research interests include service robots, pattern recognition, machine learning, and image processing.
\end{IEEEbiography}

\vfill

\end{document}